# The Planck energy-mass source as an alternative to the Big Bang

Serge F. Timashev

USPolyResearch, Ashland, PA 17921 and

Karpov Institute of Physical Chemistry, Moscow, Russia

The general theory of relativity is used to show that the total energy-mass of the visible Universe could be produced by an energy-mass source with the Planck power. The source was supposedly born at the phase of cosmic inflation and acts continuously throughout the lifetime of our Universe. The model allows one to treat dark energy as a real form of energy without using the hypothesis of anti-gravity.

98.80.Bp, 98.80.Es

For the region in the vicinity of a material object with the mass of $M$, the Schwarzschild solution to the equations of general theory of relativity (GTR) implies that the time-like component of the space-time metrics $g_0 \sim \psi(r) = \left(1 - \dfrac{r_{Sc}}{r}\right)$ tends to zero and the corresponding space-like component tends to infinity at the distance of $r = r_{Sc} = \dfrac{2GM}{c^2}$ from the object ($c$ is the light velocity in vacuum, $G$ is the gravitation constant)[1]. This result was used to show that "black" holes may exist. If we rewrite $\psi(r)$ as

$\psi(r) = \left(1 - \dfrac{w}{w_{Sc}}\right)$, where $w = \dfrac{Mc^3}{r}$ is the power of the source producing the energy of

$Mc^2$ for the time of $r/c$ and $w_{Sc} = \dfrac{Mc^3}{r_{Sc}} = \dfrac{c^5}{2G} \approx 1.8 \cdot 10^{59}$ erg/s is the power corresponding to the Schwarzschild time of $t_{Sc} = r_{Sc}/c$, then the same logic leads us to the conclusion that an energy source with the power of $w_{Sc}$ may exist.

The Planck parameters of length $l_{Pl}$, time $t_{Pl}$, and mass $m_{Pl}$ [2][3]:

$$l_{Pl} = \varsigma_l \sqrt{\dfrac{G\hbar}{c^3}} \sim 1.6 \cdot 10^{-33} \text{ sm};\quad t_{Pl} = \dfrac{l_{Pl}}{c} = \varsigma_t \sqrt{\dfrac{G\hbar}{c^5}} \sim 5.3 \cdot 10^{-44} \text{ s};$$

$$m_{Pl} = \varsigma_m \sqrt{\dfrac{\hbar c}{G}} \sim 2.2 \cdot 10^{-5} \text{ g}, \qquad (1)$$

where $\varsigma_l \sim 1$, $\varsigma_t \sim 1$, and $\varsigma_m \sim 1$ are dimensionless coefficients, were introduced by Planck in 1899 on the basis of the fundamental physical constants $c$, $G$, and the Planck constant. The Planck scales $l_{Pl}$ and $t_{Pl}$ are often used in quantum gravitation theories[2]. If we take the ratio of $\varsigma_l/\varsigma_m$ equal to 2, then the Planck power can be defined as

$w_{Pl} = \dfrac{m_{Pl} c^2}{t_{Pl}} = \dfrac{c^5}{2G} \approx 1.8 \cdot 10^{59}$ erg/s. As $w_{Pl} = w_{Sc}$, the energy source with the power of $w_{Sc}$ will further be referred to as the Planck source. It is worth noting that only the power $w_{Pl}$, which does not depend on the Planck constant $\hbar$, can be considered as a parameter of the classical GRT equations. So, the anomal changes in the space-time metrics for the Schwarzschild solution can take place not only at the distance $r_{Sc}$ from the object with mass $M$, but also near the energy source with the Planck power $w_{Pl}$.

Let us show that the source with the Planck power $w_{Pl}$ could produce the total energy-mass of the visible Universe for the period from the phase of cosmic inflation[3] up to present. The simplest form of the Einstein's equations for the expanding homogeneous and isotropic Universe can be written as[4]:

$$\left(\frac{\dot{R}}{R}\right)^2 = \frac{8\pi G}{3c^2}(\varepsilon_V + \varepsilon_{dm} + \varepsilon_b) - \frac{c^2 k}{R^2}, \quad (2)$$

$$\frac{\ddot{R}}{R} = -\frac{4\pi G}{3c^2}(\varepsilon_{dm} + \varepsilon_b - 2\varepsilon_V + 3p). \quad (3)$$

Here, $\varepsilon_V$, $\varepsilon_{dm}$, and $\varepsilon_b$ are the energy density of the physical vacuum, "dark matter", and baryonic component, respectively; $R(t)$ is the cosmological expansion (scale) factor that is independent of space coordinates; $p$ is the value of the effective pressure averaged over all galaxies and galaxy clusters; $k$ is the parameter characterizing the spacetime curvature.

We consider the case of $k = 0$ because our Universe is Euclidean[5]. Taking into account the Hubble law,

$$\dot{R} = HR, \quad (4)$$

which relates the speed of the Universe expansion to the scale factor, Eq. (2) gives the relation between the gravitational constant $G$ and the basic parameters of the Universe, Hubble's constant $H$ and total averaged energy density $\varepsilon_{tot} = \varepsilon_V + \varepsilon_{dm} + \varepsilon_b$:

$$G = \frac{3c^2}{8\pi(\varepsilon_V + \varepsilon_{dm} + \varepsilon_b)} \cdot \frac{\dot{R}^2}{R^2} = \frac{3c^2 H^2}{8\pi \varepsilon_{tot}}. \quad (5)$$

Assuming that the cosmological parameters $c$, $H$, $\varepsilon_V$, $\varepsilon_{dm}$, and $\varepsilon_b$ do not depend on time, Eqs. (2) and (3) lead to

$$\frac{\ddot{R}}{R} = H^2, \quad p = -(\varepsilon_{dm} + \varepsilon_b) \equiv -\varepsilon_m. \tag{6}$$

According to the present estimates[6], $\varepsilon_{tot} \approx 0.9 \cdot 10^{-8}$ erg/sm$^3$. At the same time, the relative contributions of the vacuum energy density $\varepsilon_V/\varepsilon_{tot}$, "dark matter" $\varepsilon_{dm}/\varepsilon_{tot}$, and baryonic component $\varepsilon_b/\varepsilon_{tot}$ are equal to 0.73, 0.23, and 0.04, respectively. For our epoch, $H = 73$ km/(s·Mpc) $\approx 2.36 \cdot 10^{-18}$ 1/s [6]. It is necessary to add that $H^{-1} = T_0$, where $T_0$ is the age of the Universe ($T_0 \approx 13.7$ billion years), and $R_0 = c/H \sim 10^{28}$ cm is the estimated value for the Universe "radius".

The second expression in Eq. (6) is usually considered as the "state equation" for the scalar field φ and is assumed to be related to the physical vacuum in the modern models of the Universe expansion[7]. According to the models, the expansion of the Universe does not reduce the scalar-field energy because the pressure of the vacuum component is negative and the energy-momentum tensor of the scalar field is chosen to be proportional to the metric tensor. In this case, the decrease in the energy density of the scalar field associated with the Universe expansion is fully compensated by the "negative" work performed by the expanding volume element. Theoretical understanding of this component ("dark energy"), which is driving the accelerated expansion of the Universe, is one of the fundamental problems in modern cosmology[8].

The introduction of the concept of a continuously acting Plank power source allows one to give a different interpretation to Eq. (6) without treating the physical vacuum as an antigravitational substance. Taking into account Eq. (5), it is easy to show that the Planck source, which has continuously been "on" throughout the lifetime of our Universe $T_0$, could produce the total energy $E_{tot}$ of the Universe:

$$w_{Pl}T_0 = \frac{c^5}{2G} \cdot \frac{1}{H} = \frac{4\pi c^3}{3H^3}\varepsilon_{tot} = \frac{4}{3}\pi R_0^3 \varepsilon_{tot} = \varepsilon_{tot}V_0 \equiv E_{tot}, \tag{7}$$

where $V_0 = 4\pi R_0^3/3$ is the Universe volume.

In contrast to the standard Big Bang hypothesis, this estimate suggests a model of a continuously acting energy-mass source that was "turned on" after the phase of cosmic inflation, which was the starting and most "energized" (based on the space coverage for such a short time period) stage of Universe formation. In this case, the expansion of the Universe should be considered as a piston-like pushing of galaxy clusters in the outward direction, which are virtually "frozen" (the effect of dark matter) into the energy-saturated medium of the physical vacuum. So, there is no need to assume that dark energy works opposite gravity and is just speeding up the expansion of the Universe.

One may estimate the effective pressure $p_{eff}$ of such a piston based on the work required to form new space by assuming that the radius of the Universe increases under stationary conditions by the value of Δ$R$ for the time of Δ$t$. Taking into account Eqs. (4) and (7), one obtains:

$$w_{Pl}\Delta t = p_{eff}\Delta V\Big|_{R_0} = p_{eff}\frac{\Delta V}{\Delta R}\cdot\frac{\Delta R}{\Delta t}\Big|_{R_0}\Delta t = p_{eff}\cdot 4\pi R_0^3 H\Delta t, \qquad (9)$$

which leads to

$$p_{eff} = \frac{c^5}{8\pi R_0^3 HG} = \frac{1}{3}\varepsilon_{tot}. \qquad (10)$$

The estimate of $p_{eff} \approx 0.33\varepsilon_{tot}$ differs by only 20% from the absolute value $|p| \approx 0.27\varepsilon_{tot}$ determined with Eq. (6). This insignificant difference can be attributed to the fact that the pressure $p$ in Eq. (3) is the value averaged over all galaxies and galaxy clusters. The low value of this difference supports the hypothesis that the energy-mass Plank source could be driving the Universe expansion. In this case, expression (6) should no longer be considered as the "state equation" because it characterizes the averaged value of the pressure that is "pushing the space out".

So, there is no need to assume that dark energy works opposite gravity and is just speeding up the expansion of the Universe. In this case, dark energy may be considered as a material substance that takes place in real phenomena, such as the Casimir effect[9], Lamb shift[10], registered noise spectrum in Josephson junctions[11]. According to the equality of $w_{Sc} = w_{Pl}$, the ratio of the generated "mass" $M$ of this source to its radius $r$ can be estimated as $M/r \approx 0.67\cdot 10^{28}$ g/cm. To illustrate the scale of this value, let us note that $M_S/R_S \approx 2.86 \cdot 10^{22}$ g/cm, where $M_S$ and $R_S$ are the mass and radius of the Sun, respectively. It can be suggested that dark matter, produced by the Plank source, together with dark energy and baryonic component determines the formation of galaxies and their clusters. The dark matter is localized in these galaxies and clusters, and forms single material blocks with the baryonic component: "gravitation polarons" or "gravitation lenses".

The basic question is how to introduce the Planck source in the models that describe the dynamics of the Universe. The GRT equations are written for the dynamics of the $R(t)$ metric and do not depend on space coordinates. As these equations are based on averaging for all galaxies and galaxy clusters, the effects of the Planck source in Eqs. (2)-(3) are implicit. To introduce the Planck source in the cosmological models, one needs to add to the GRT system a system of equations that describe the space dynamics of galaxy clusters and dark matter in the physical vacuum. These equations must include the mechanism of energy and momentum transfer from the Planck source to the Universe treated as a single system. This mechanism must lead to the realization of the cosmological principle stating that large-scale Universe is homogeneous and isotropic.

The verifiable conclusions of the Planck source model are very similar to those for the scalar-field model if one does not consider the specific mechanism of energy transfer from the source to the local areas of the Universe. On the other had, let us note that the proposed model is conceptually close to Hoyle's steady state theories.[12]

[1] L.D. Landau, and E.M. Lifshitz, *The Classic Theory of Fields*, **4th Ed.** (Butterworth-Heinemann, Oxford, 1975).

[2] L. Smolin, Sci. Amer., 56 (Jan. 2004).

[3] B. Greene, *The Elegant Universe* (Vintage Books, New York, 1999).

[4] H.V. Klapdor-Kleingrothaus, K. Zuber, *Teilchenastrophysik* (B.G. Teubner GmbH, Stuttgart, 1997).

[5] V. Springer, S.D.M. White, and A. Jenkins, Nature (London) **435**, 629 (2005).

[6] T. Reichhardt, Nature (London) **421**, 777 (2003).

[7] A. Linde, *Particle Physics and Inflationary Cosmology* (CRC Press, Boca Raton, Fl, 1990)

[8] T. Padmanabhan, 29 International Cosmic Ray Conference, Pune, India **10**, 47, (2005).

[9] G. Plunien, B. Muller, and W. Greiner, Phys. Rep. **134**, 87 (1986).

[10] G.W. Erickson, Phys. Rev. Lett. **27**, 780 (1971).

[11] C. Beck and M.C. Mackey, Fluct. Noise Lett. **7**, C27 (2007).

[12] F. Hoyle, G. Burbidge and J. V. Narlikar, Astrophys. J., **410**, 437 (1993).